\documentstyle[11pt,epsfig,fullpage,citesort]{article}

\makeatletter

\def\@sect#1#2#3#4#5#6[#7]#8{\ifnum #2>\c@secnumdepth
     \def\@svsec{}\else 
     \refstepcounter{#1}\edef\@svsec{\csname the#1\endcsname.\hskip 1em }\fi
     \@tempskipa #5\relax
      \ifdim \@tempskipa>\z@ 
        \begingroup #6\relax
          \@hangfrom{\hskip #3\relax\@svsec}{\interlinepenalty \@M #8\par}
        \endgroup
       \csname #1mark\endcsname{#7}\addcontentsline
         {toc}{#1}{\ifnum #2>\c@secnumdepth \else
                      \protect\numberline{\csname the#1\endcsname}\fi
                    #7}\else
        \def\@svsechd{#6\hskip #3\@svsec #8\csname #1mark\endcsname
                      {#7}\addcontentsline
                           {toc}{#1}{\ifnum #2>\c@secnumdepth \else
                             \protect\numberline{\csname the#1\endcsname}\fi
                       #7}}\fi
     \@xsect{#5}}

\renewcommand{\section}{\setcounter{equation}{0} \@startsection {section}{1}
   {\z@}{-3.5ex plus -1ex minus -.2ex}{2.3ex plus .2ex}{\Large\bf}}
\newcommand{\fcaption}[1]{
        \refstepcounter{figure}
        \setbox\@tempboxa = \hbox{\small Fig.~\thefigure. #1}
        \ifdim \wd\@tempboxa > 6in
           {\begin{center}
        \parbox{6in}{\small\baselineskip=12pt Fig.~\thefigure. #1}
            \end{center}}
        \else
             {\begin{center}
             {\small Fig.~\thefigure. #1}
              \end{center}}
        \fi}
\newcommand{\tcaption}[1]{
        \refstepcounter{table}
        \setbox\@tempboxa = \hbox{\small Table~\thetable. #1}
        \ifdim \wd\@tempboxa > 5.4in
           {\begin{center}
        \parbox{5.4in}{\small\baselineskip=12pt Table~\thetable. #1}
            \end{center}}
        \else
             {\begin{center}
             {\small Table~\thetable. #1}
              \end{center}}
        \fi}

\def\PRD{{\em Phys. Rev.} D}
\def\ZPC{{\em Z. Phys.} C}

\def\ra{\rightarrow}

\def\be{\begin{equation}}
\def\ee{\end{equation}}
\def\bea{\begin{eqnarray}}
\def\eea{\end{eqnarray}}

\newcommand{\GG}{\mbox{$\gamma\gamma$}}

\newcommand{\ggam}{\mbox{$\gamma\gamma\,$}}

\def\lsim{\mathrel{\raise.3ex\hbox{$<$\kern-.75em\lower1ex\hbox{$\sim$}}}}
\def\gsim{\mathrel{\raise.3ex\hbox{$>$\kern-.75em\lower1ex\hbox{$\sim$}}}}
\def\la{\langle}
\def\ra{\rangle}

\def\a{\alpha}

\def\g{\gamma}

\title{Electroweak gauge boson production at $\g\g$ collider
\footnote{Talk presented at the {\it International Workshop on
Linac--Ring Type $ep$ and $\gamma p$ Colliders}, 9--11 April 1997,
Ankara, Turkey}}

\author{G. Jikia\thanks{Alexander von Humboldt Fellow;
	e-mail: jikia@phyv4.physik.uni-freiburg.de} \\
[1ex]{\small\it Albert--Ludwigs--Universit\"{a}t Freiburg,
           Fakult\"{a}t f\"{u}r Physik}\\
      {\small\em Hermann--Herder Str.3, D-79104 Freiburg, Germany} \\
	{\small and} \\
{\small\it Institute for High Energy Physics, Protvino}\\
     {\small\it Moscow Region 142284, Russian Federation} }

\date{}

\begin{document}

\maketitle

\section{Introduction}

Linear colliders offer unique opportunities to study high energy
photon-photon collisions obtained using the process of Compton
backscattering of laser light off electron beams from the linear
collider \cite{Telnov97}. This option is included now in conceptual
design reports of the NLC, JLC and TESLA/SBLC projects of $e^+e^-$
linear collider \cite{NLC,CDR}.  The expected physics at the Photon
Linear Collider (PLC) is very rich and complementary to that in
$e^+e^-$ collisions. In particular PLC will be especially attractive
tool in probing the the electroweak symmetry breaking sector via
precision measurements of anomalous $W$ self couplings.
In this paper a short survey of the most important processes of electroweak
gauge boson production in photon-photon collisions is given.

\section{$\g\g\to W^+W^-$ cross sections and quantum ${\cal O}(\a)$ 
corrections}

The reaction $\g\g\to W^+W^-$ would be the dominant source of the
$W^+W^-$ pairs at future linear colliders, provided that photon-photon
collider option will be realized. The Born cross section of $W^+W^-$
pair production in photon-photon collisions in the scattering angle
interval $10^\circ < \theta^\pm < 170^\circ$ is 61~pb at
$\sqrt{s_{\g\g}}=500$~GeV and 37~pb at 1~TeV.  Corresponding cross
sections of $W^+W^-$ pair production in $e^+e^-$ collisions are an
order of magnitude smaller: 6.6~pb at 500~GeV and 2.5~pb at 1~TeV.
With more than a million $WW$ pairs per year a photon-photon collider
can be really considered as a $W$-factory and an ideal place to
conduct precision tests on the anomalous triple and quartic couplings
of the $W$ bosons.

With the natural order of magnitude on anomalous couplings one needs
to know the ${\cal SM}$ cross sections with a precision better than
1\% to extract these small numbers. From a theoretical point of view
this calls for the very careful analysis of at least $\cal{O}(\alpha)$
corrections to the cross section of $W^+W^-$ pair production in $\g\g$
collisions, which were recently calculated including virtual
corrections \cite{DennerDittmaierSchusterAAWW} and including complete
$\cal{O}(\alpha)$ corrections with  account of both virtual
one-loop corrections and real photon and $Z$-boson emission
\cite{JikiaAAWW}.

\begin{figure}
\setlength{\unitlength}{1in}
\begin{picture}(6,3.5)
\put(-.15,0){\epsfig{file=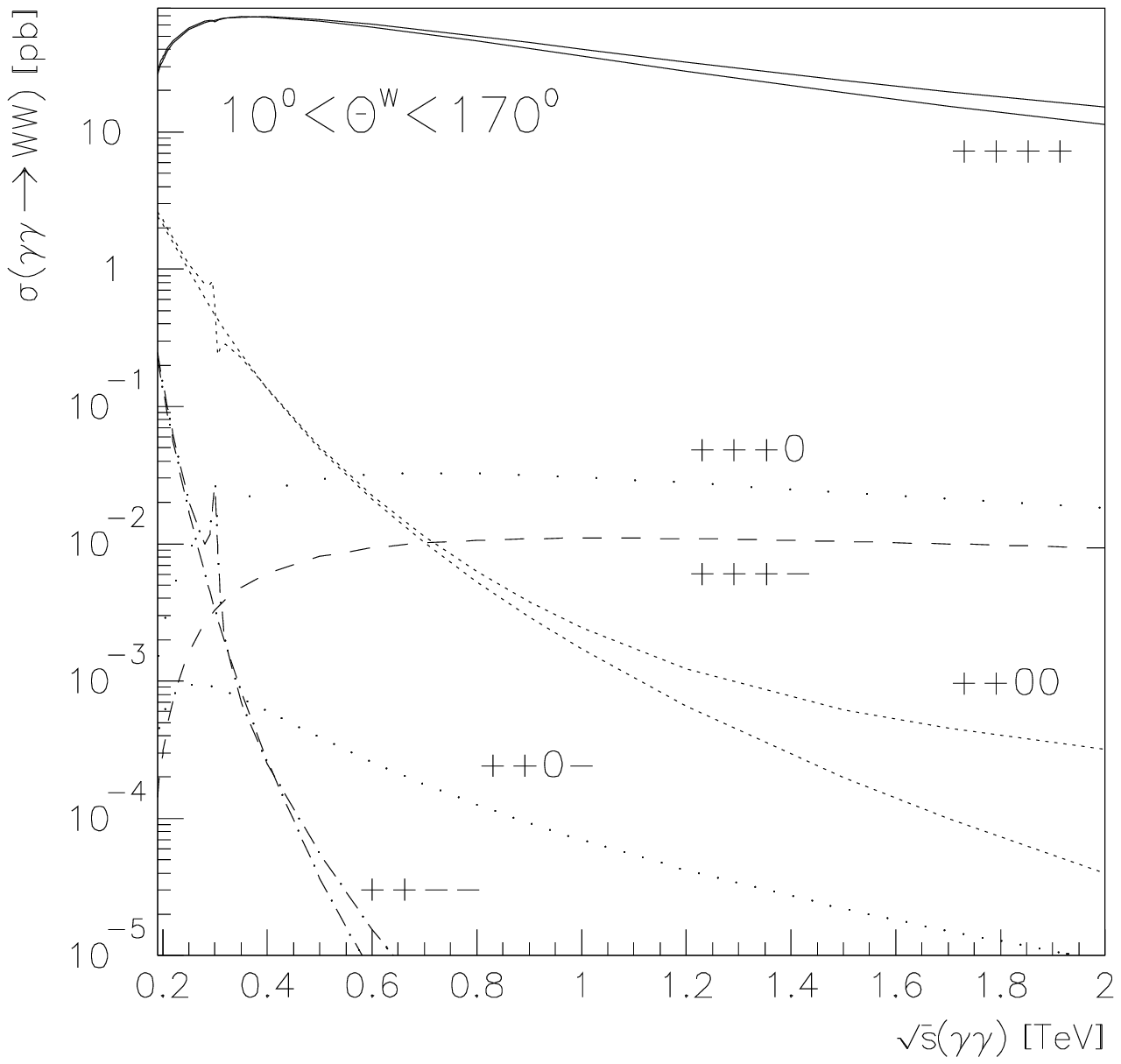,width=3.2in,height=3.5in}}
\put(2.85,0){\epsfig{file=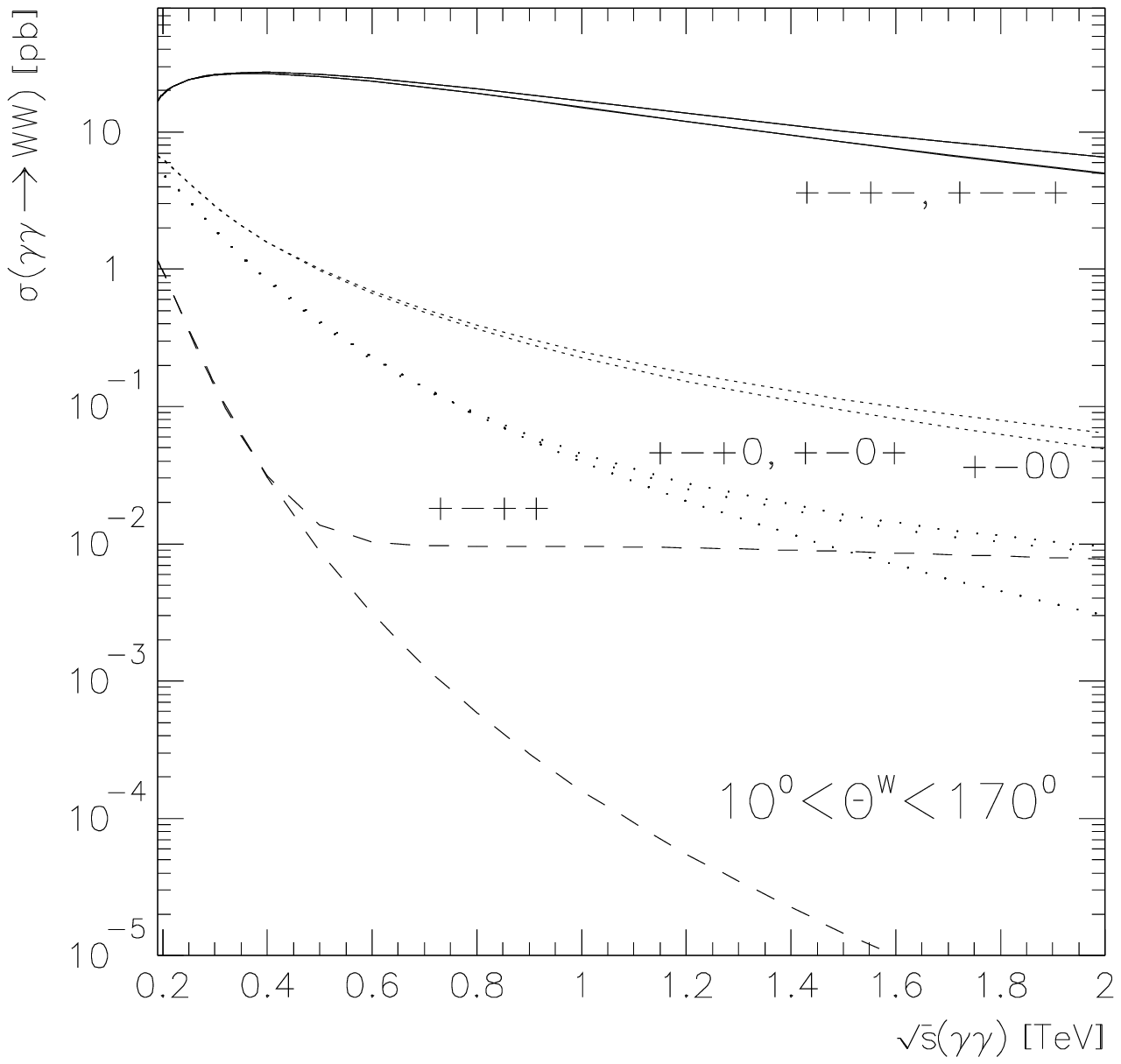,width=3.2in,height=3.5in}}
\end{picture}
\fcaption{Total cross sections of $WW(\g)$ production for various
polarizations.  Born and corrected cross sections are shown. The
curves nearest to the helicity notations represent the corrected cross
sections.}
\end{figure}

Figure~1 shows total cross section of $WW$ pair production summed over
$WW$ and $WW\g$ final states and integrated over $W^\pm$ scattering
angles in the interval $10^\circ<\theta^\pm<170^\circ$ as a function
of energy for various polarizations \cite{JikiaAAWW}.  The bulk of the
cross section originates from transverse $W_TW_T$ pair production.
Transverse $W$'s are produced predominantly in the forward/backward
direction and the helicity conserving amplitudes are dominating. Cross
sections integrated over the whole phase space are non-decreasing with
energy.  For a finite angular cutoff they do decrease as $1/s$, but
still they are much larger than suppressed cross sections. For the
dominating $++++$, $+-+-$, $+--+$ helicity configurations corrections
are negative and they rise with energy ranging from $-3\%$ at 500~GeV
to $-25\%$ at 2~TeV.

\begin{table}
\tcaption{Total unpolarized Born cross sections and relative
corrections for various intervals of $W^\pm$ scattering
angles. Corrections originating from real hard photon
($\omega_\g>k_c=0.1$~GeV) and $Z$-boson emission as well as IR-finite
sum of soft photon and virtual boson contributions, fermion virtual
corrections and total corrections are given separately.}
\begin{center}
\begin{center}
$\sqrt{s} = 300$~GeV
\end{center}
\begin{tabular}{|c|c|c|c|c|c|c|}\hline\hline
$\theta_{W^\pm}$, ${}^\circ$ & $\sigma^{Born}$, $pb$ & $\delta^{hard}$, \% &
$\delta^{Z}$, \% & $\delta^{soft+bose}$, \% & $\delta^{fermi}$, \% 
&$\delta^{tot}$, \% \\ \hline
$  0^\circ < \theta < 180^\circ$ &  70.22     &  4.15    
& 2.64$\cdot 10^{-2}$& $-$7.09    & 0.327    & $-$1.37    
 \\
$ 10^\circ < \theta < 170^\circ$ &  64.46     &  4.11    
& 2.74$\cdot 10^{-2}$& $-$7.31    & 0.257    & $-$1.59    
 \\
$ 30^\circ < \theta < 150^\circ$ &  38.15     &  4.09    
& 3.27$\cdot 10^{-2}$& $-$8.62    & $-$0.123    & $-$2.67    
 \\
$ 60^\circ < \theta < 120^\circ$ &  12.96     &  4.02    
& 2.94$\cdot 10^{-2}$& $-$10.7    & $-$0.415    & $-$3.75    
\\
\hline\hline   
\end{tabular}
\vspace{.5cm}

\begin{center}
$\sqrt{s} = 500$~GeV
\end{center}
\begin{tabular}{|c|c|c|c|c|c|c|}\hline\hline
$\theta_{W^\pm}$, ${}^\circ$ & $\sigma^{Born}$, $pb$ & $\delta^{hard}$, \% &
$\delta^{Z}$, \% & $\delta^{soft+bose}$, \% & $\delta^{fermi}$, \% 
&$\delta^{tot}$, \% \\ \hline
$  0^\circ < \theta < 180^\circ$ &  77.50     &  7.96    & 0.468    
& $-$10.1    & 9.04$\cdot 10^{-2}$& $-$1.63    
 \\
$ 10^\circ < \theta < 170^\circ$ &  60.71     &  7.89    & 0.541    
& $-$10.7    & $-$0.242    & $-$2.52    
 \\
$ 30^\circ < \theta < 150^\circ$ &  21.85     &  8.05    & 0.817    
& $-$13.0    & $-$1.34    & $-$5.50    
 \\
$ 60^\circ < \theta < 120^\circ$ &  5.681     &  8.02    & 0.789    
& $-$14.8    & $-$2.13    & $-$8.12
\\    
\hline\hline   
\end{tabular}
\vspace{.5cm}

\begin{center}
$\sqrt{s} = 1000$~GeV
\end{center}
\begin{tabular}{|c|c|c|c|c|c|c|}\hline\hline
$\theta_{W^\pm}$, ${}^\circ$ & $\sigma^{Born}$, $pb$ & $\delta^{hard}$, \% &
$\delta^{Z}$, \% & $\delta^{soft+bose}$, \% & $\delta^{fermi}$, \% 
&$\delta^{tot}$, \% \\ \hline
$  0^\circ < \theta < 180^\circ$ &  79.99     &  13.3    &  1.55    
& $-$18.7    & $-$5.51$\cdot 10^{-2}$& $-$3.89    
 \\
$ 10^\circ < \theta < 170^\circ$ &  37.04     &  13.4    &  2.39    
& $-$22.6    & $-$1.28    & $-$8.10    
 \\
$ 30^\circ < \theta < 150^\circ$ &  6.924     &  14.2    &  3.96    
& $-$32.1    & $-$3.80    & $-$17.8    
 \\
$ 60^\circ < \theta < 120^\circ$ &  1.542     &  14.2    &  3.88    
& $-$37.1    & $-$5.13    & $-$24.1    
 \\
\hline\hline   
\end{tabular}
\vspace{.5cm}

\begin{center}
$\sqrt{s} = 2000$~GeV
\end{center}
\begin{tabular}{|c|c|c|c|c|c|c|}\hline\hline
$\theta_{W^\pm}$, ${}^\circ$ & $\sigma^{Born}$, $pb$ & $\delta^{hard}$, \% &
$\delta^{Z}$, \% & $\delta^{soft+bose}$, \% & $\delta^{fermi}$, \% 
&$\delta^{tot}$, \% \\ \hline
$  0^\circ < \theta < 180^\circ$ &  80.53     &  19.0    &  2.91    
& $-$27.2    & $-$7.45$\cdot 10^{-2}$& $-$5.33    
 \\
$ 10^\circ < \theta < 170^\circ$ &  14.14     &  20.1    &  6.38    
& $-$41.6    & $-$2.99    & $-$18.1    
 \\
$ 30^\circ < \theta < 150^\circ$ &  1.848     &  21.5    &  9.77    
& $-$60.1    & $-$6.54    & $-$35.4    
 \\
$ 60^\circ < \theta < 120^\circ$ & 0.3936     &  21.6    &  9.60    
& $-$67.6    & $-$8.04    & $-$44.5    
\\
\hline\hline   
\end{tabular}
\end{center}
\end{table}

In Table~1 Born cross sections and relative corrections are given for
several intervals of $W^\pm$ scattering angles \cite{JikiaAAWW}. At
high energies large cancellations occur between negative virtual
corrections and positive corrections corresponding to real photon or
$Z$-boson emission.  Consequently, although the correction originating
from the $WWZ$ production is completely negligible at
$\sqrt{s_{\g\g}}=0.3$~TeV, it is of the same order of magnitude as
hard photon correction at 2~TeV.  Although at $300\div 500$~GeV
corrections are quite small ranging from $-1.3\%$ to $-8\%$, depending
on angular cuts, at TeV energies the value of radiative corrections in
the central region of $W^+W^-$ production become quite large, so that
corrections in the region $60^\circ < \theta < 120^\circ$ are $6\div
8$ times larger than the corrections to the total cross section at
$1\div 2$~TeV. They range from $-24\%$ to $-45\%$. Thus if precision
measurements are to be made at TeV energy, more careful theoretical
analysis could be needed in order to reliably predict the value of the
cross section in the central region where the value of the cross
section is the most sensitive to the $W$ anomalous couplings.

\section{$\g\g\to ZZ$ production}

$Z$-pair production in photon-photon collisions plays a special role
due to the possibility to observe the Higgs signal in \GG\ collisions
for the Higgs bosons heavier that $2M_Z$ in $ZZ$ decay mode
\cite{GunionHaber,BBC} if one of the $Z$'s is required to decay to
$l^+l^-$ to suppress the huge tree-level $\gamma\gamma\to W^+W^-$
continuum background. However, even though there is no tree-level $ZZ$
continuum background, such a background due to the reaction
$\gamma\gamma\to ZZ$ does arise at the one-loop level in the
electroweak theory
\cite{JikiaAAZZ,BergerAAZZ,DicusKaoAAZZ,VeltmanAAZZ} which makes the
Higgs observation in the $ZZ$ mode impossible for $m_h\gsim (350\div
400)$~GeV. It was found that for $185\lsim m_h\lsim 300$~GeV the $ZZ$
mode will provide a 10-20\% determination of the quantity
$\Gamma(h\to\gamma\gamma)\cdot BR(h\to ZZ)$. 

\begin{figure}
\setlength{\unitlength}{1in}
\begin{picture}(6,3.5)
\put(-.15,0){\epsfig{file=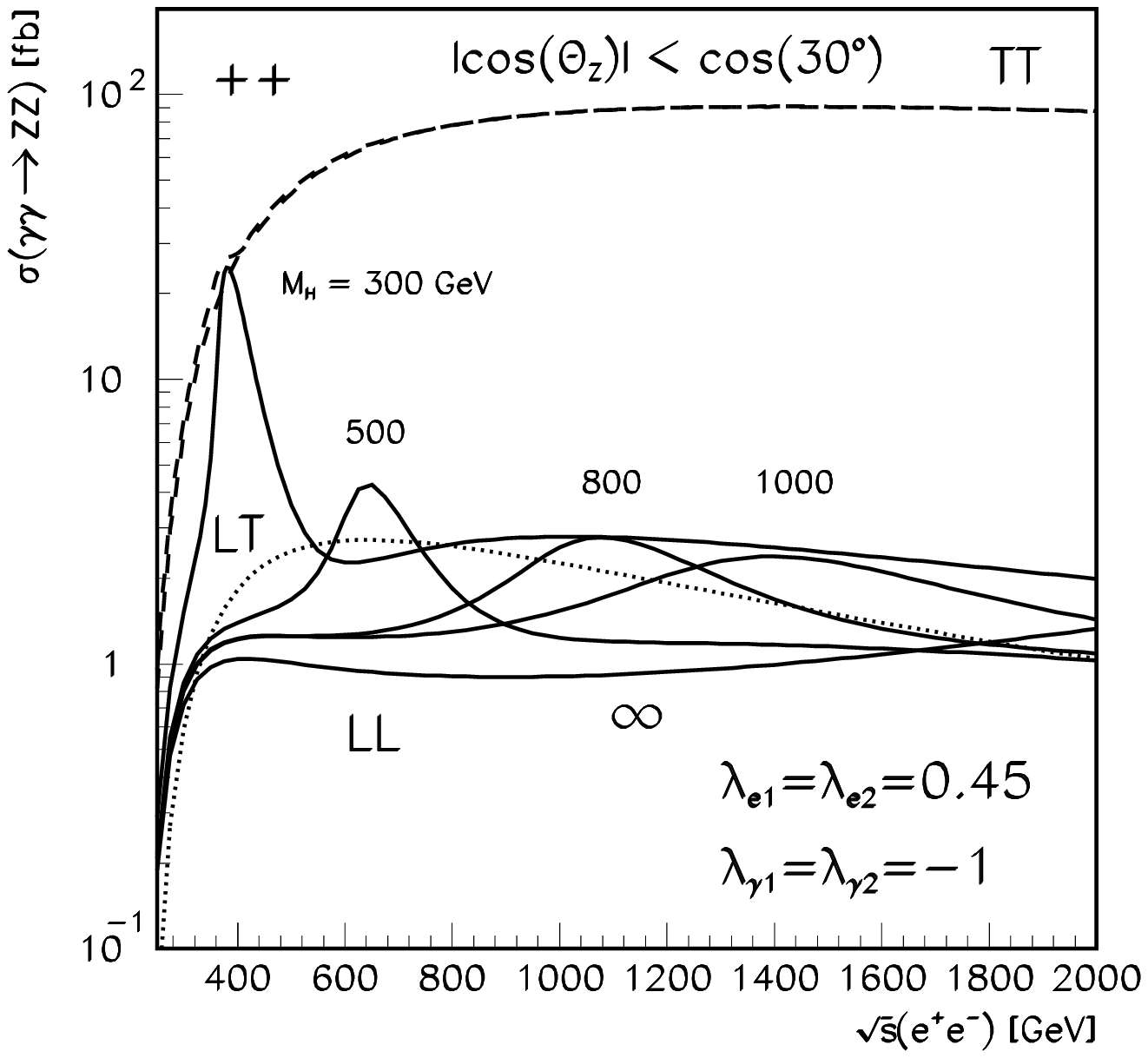,width=3.2in,height=3.5in}}
\put(2.85,0){\epsfig{file=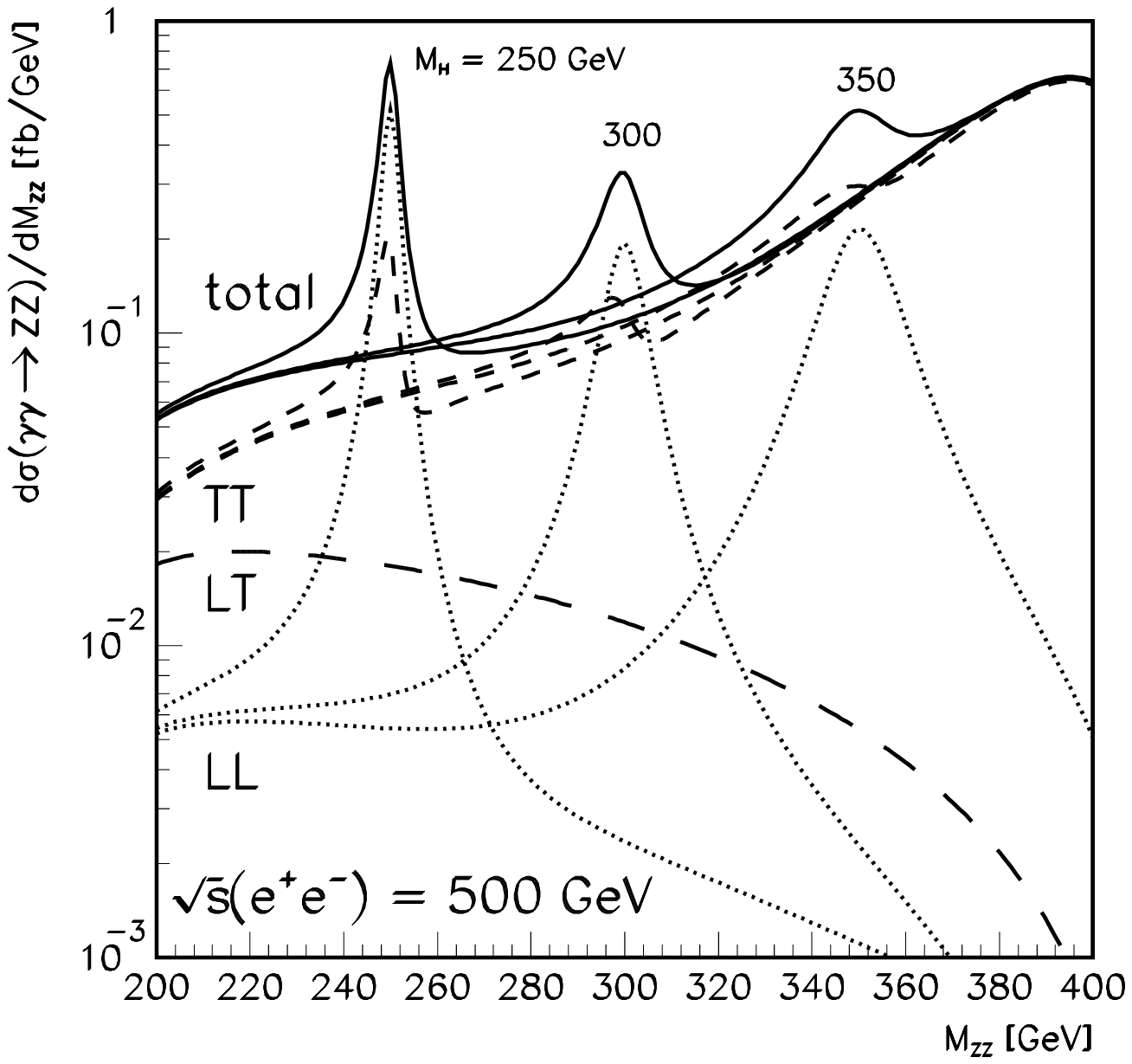,width=3.2in,height=3.5in}}
\end{picture}

\fcaption{(a) Cross section of the $ZZ$ pair production in polarized
$\g\g$ collisions versus c.m.s. energy of the $e^+e^-$ collisions
computed taking into account photon spectrum of the backscattered
laser beams. Both $Z$-bosons have $|\cos \theta_Z|<\cos
30^{\mbox{o}}$. Curves for $Z_LZ_L$ (solid line), $Z_TZ_T$ (dashed
line) and $Z_LZ_T$ (dotted line) production are shown. Different
curves for longitudinal $Z_LZ_L$ pair production correspond to Higgs
boson masses of 300, 500, 800, 1000~GeV and infinity.\\
(b) The invariant mass, $M_{ZZ}$, distribution of $Z$-bosons for
$\g\g\to ZZ$ in photon-photon collisions at
$\sqrt{s_{e^+e^-}}=500$~GeV and $m_H=250$, 300 and 350~GeV. Curves for
$Z_LZ_L$ (dotted line), $Z_TZ_T$ (dashed line) and $Z_LZ_T$ (long
dashed line) production are shown in addition to the sum over all
polarizations of the $Z$-boson (solid line).}
\end{figure}

In Fig.~2 the cross section of the $ZZ$ pair production and invariant
mass distribution at the PLC are shown \cite{JikiaAAZZ}. With the
polarizations of the initial electron (positron) and laser beams shown,
the photon-photon energy spectrum peaks just below the highest allowed
photon-photon energy and colliding photons are produced mainly with
equal mean helicities $\la\xi_1\xi_2\ra\sim 1$. As for the case of $W$
pair production, at high energies the cross section is dominated by
the transversely polarized $Z_TZ_T$ pair production.  As it was
already mentioned, while clear Higgs boson peaks are observable at
$\sqrt{s_{e^+e^-}}=500$ GeV for $m_H=250$ and 300 GeV in Fig.~2b, a
background from transverse $Z_TZ_T$ pair production makes the
observation of heavier Higgs signal quite problematic.

\begin{figure}
\setlength{\unitlength}{1in}
\begin{picture}(6,3.5)
\put(-.15,0){\epsfig{file=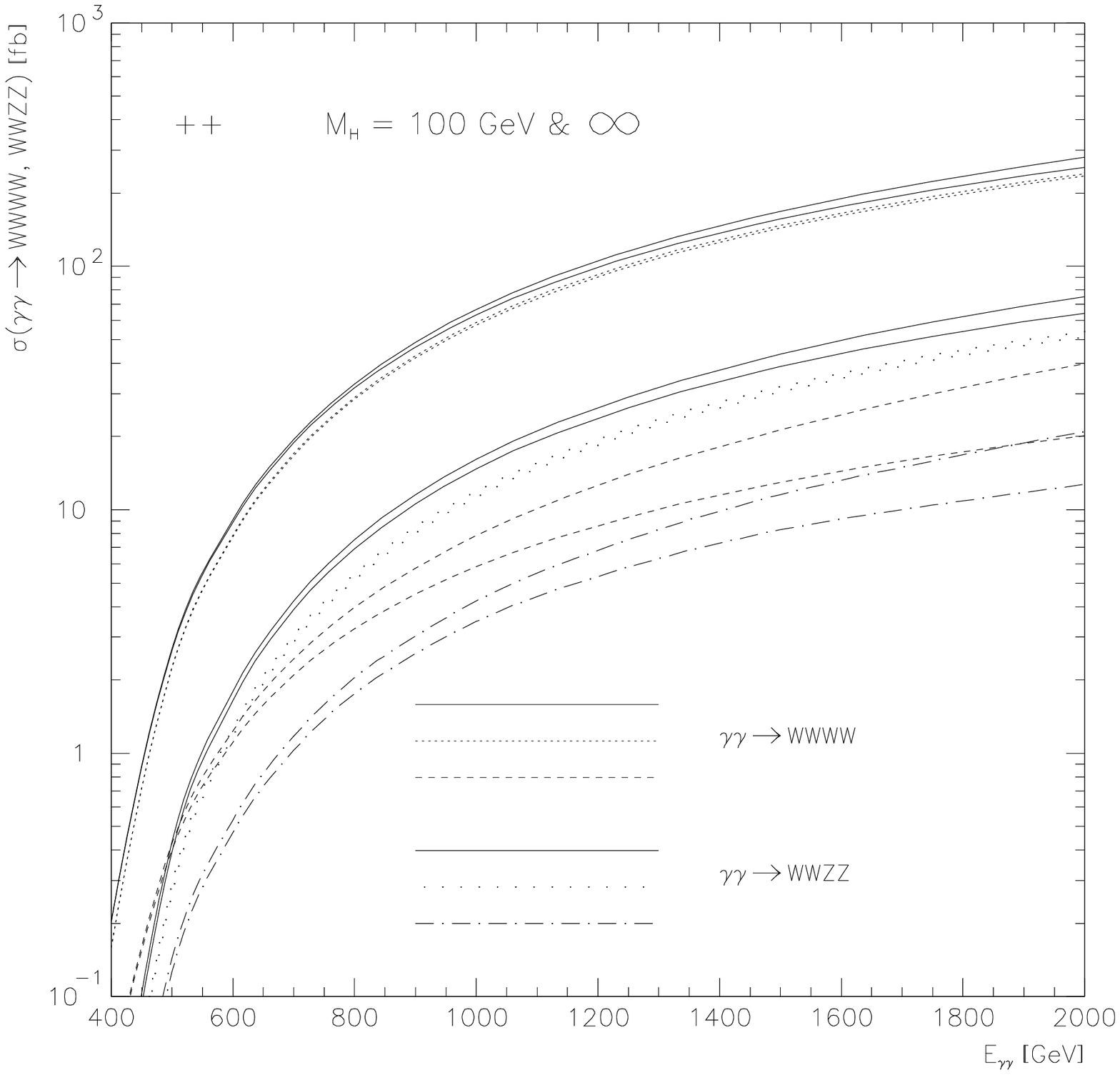,width=3.2in,height=3.5in}}
\put(2.85,0){\epsfig{file=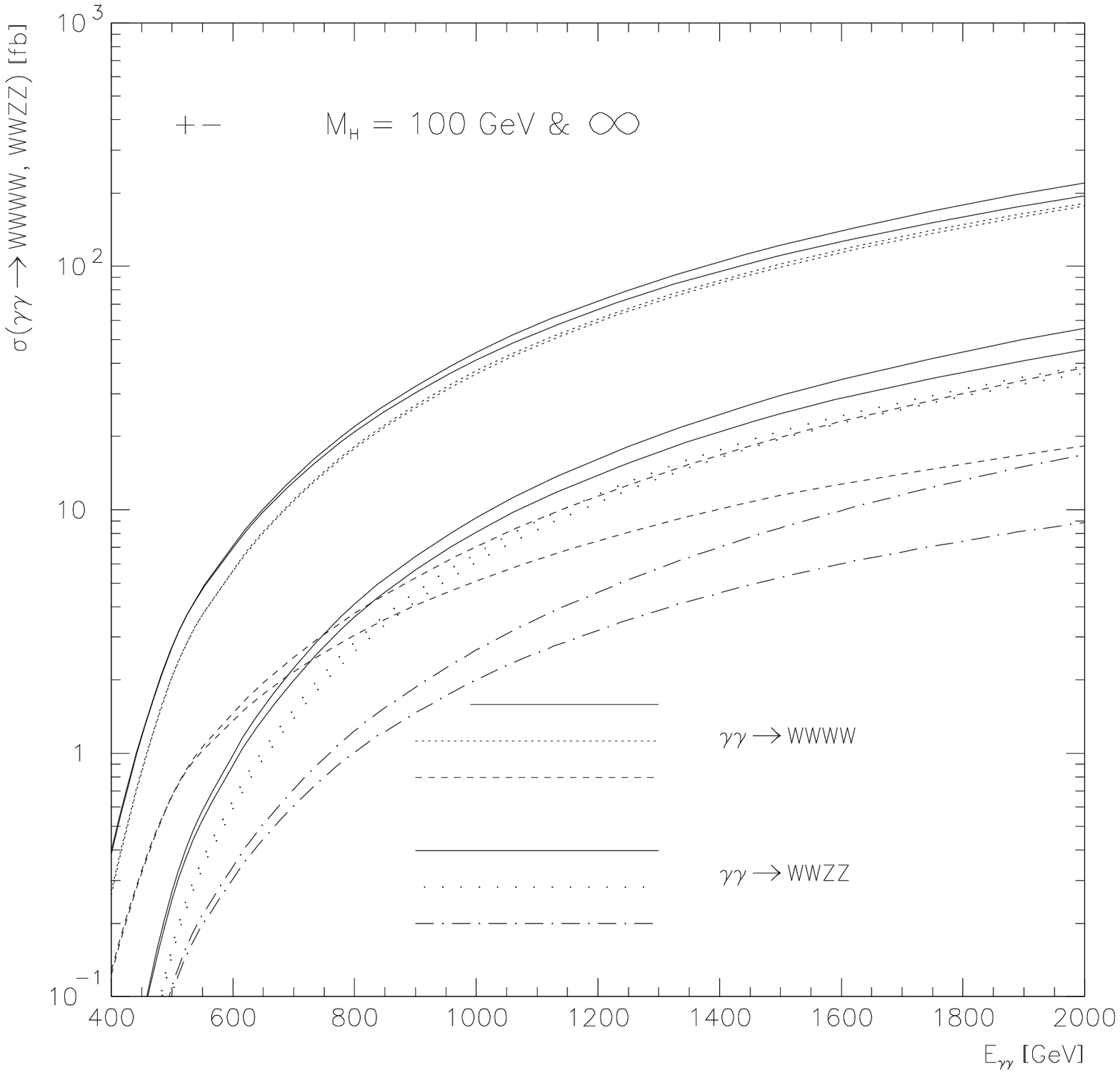,width=3.2in,height=3.5in}}
\end{picture}
\fcaption{Comparison between the cross sections for $m_H=100$~GeV and
$m_H=\infty$ for equal and opposite helicities of the initial
photons.  For the reaction $\g\g\to WWWW$ the following cross sections
are shown: total cross section (solid line), the $TTTT+TTTL$ cross
sections (dotted line), the sum of cross sections with at least two
longitudinal final $W$'s (dashed line). For the reaction $\g\g\to
WWZZ$ corresponding cross sections are denoted by solid, dotted and
dash-dotted lines.}
\end{figure}

\section{$W^+W^-\to W^+W^-$, $ZZ$ scattering}

At center-of-mass energy above 1~TeV the effective $W$ luminosity
becomes substantial enough to allow for the study of $W^+W^-\to
W^+W^-$, $ZZ$ scattering in the reactions $\ggam\to WWWW$, $WWZZ$,
when each incoming photon turns into a virtual $WW$ pair, followed by
the scattering of one $W$ from each such pair to form $WW$ or
$ZZ$. 

It was found \cite{JikiaWWWW,CheungWWWW}  that a signal of SM
Higgs boson with $m_h$ up to 700~GeV (1~TeV) could be probed in these
processes at 1.5~TeV (2~TeV) PLC, assuming integrated luminosity of
200~fb$^{-1}$ (300~fb$^{-1}$). However even larger luminosity is
needed in order to extract the signal of enhanced $W_LW_L$ production
in models of electroweak symmetry breaking without Higgs boson
\cite{JikiaWWWW}. The main problem is again large background from
transverse $W_TW_TW_TW_T$, $W_TW_TZ_TZ_T$ production.

Event rates as well as signal/background ratio and the statistical
significance corresponding to various values of the Higgs boson mass
and cosine of the dead cone angle $z_0$ are given in Table~2 for total
energies of 1.5 and 2~TeV \cite{JikiaWWWW}.  The value of integrated
luminosity of 200~fb$^{-1}$ is assumed and branching ratio of 50\% for
hadronic decays of $WW$, $ZZ$ pairs is included. At $\sqrt{s}=1.5$~TeV
we require that the invariant mass $M_{34}$ of central pair $WW$, $ZZ$
lie in the interval 400~GeV$<M_{34}<$~600~GeV for $m_H=500$~GeV and
500~GeV$<M_{34}<$~800~GeV for $m_H=700$~GeV. For $m_H=1$~TeV and
$\sqrt{s}=2$~TeV 450~GeV$<M_{34}<$~1.1~TeV.

\begin{table}[htb]
\tcaption{Event rates for signal ($S$) and background ($B$) summed
over $WWWW$ and $WWZZ$ final states as well as signal/background ratio
and statistical significance. }
\begin{center}
\begin{tabular}{|c|c|cccc||cccc|} \hline\hline
\multicolumn{2}{|c|}{}
&\multicolumn{4}{|c||}{$z_0=\cos(10^\circ)$}
&\multicolumn{4}{|c|}{$z_0=\cos(5^\circ)$}\\ \hline
$\sqrt{s_{e^+e^-}}$, TeV & $m_H$, GeV & $S$ & $B$ & $S/B$ & $S/\sqrt{B}$ 
& $S$ & $B$ & $S/B$ & $S/\sqrt{B}$  \\ \hline
1.5 & 500  & 84 & 34 & 2.5  & 14  & 218 & 56 & 3.9 & 29 \\
    & 700  & 24 & 23 & 1.0  & 5.0 &  53 & 37 & 1.4 & 8.7 \\ \hline
2   & 1000 & 14 & 21 & 0.67 & 3.0 &  74 & 59 & 1.3 & 9.6 \\ \hline\hline
\end{tabular}
\end{center}
\end{table}

As one can see from Fig.~3 the contribution from two longitudinal weak
bosons $TTLL$ and $TLTL$, which are sensitive to heavy Higgs boson
contribution, are about an order of magnitude smaller than that for
$TTTT$ production \cite{JikiaWWWW}. So, for the total cross sections
one should expect 10$\%$ signal-to-background ratio.

\section{$\g\g\to \g\g$, $\g Z$}

Neutral gauge boson $\g\g$, $\g Z$, $ZZ$ pair production processes in
photon--photon fusion represent special interest because these
processes are absent at the classical level and are generated at the
one-loop level due to quantum corrections. The collision of high
energy, high intensity photon beams at the Photon Collider would
provide novel opportunities for such processes.  The distinctive
feature of the electroweak vector boson loops contribution is that it
leads to the differential cross sections behaving as $d\sigma/dt\propto
1/t^2$ in the high energy limit and, hence, to a nondecreasing with
energy total cross sections.

The total cross sections of $\g\g$, $\g Z$ pair production are shown
in Figs.~4, 5, respectively. $W$ loop contribution dominates at
photon-photon collision energies above 250~GeV.

\begin{figure}
\setlength{\unitlength}{1in}
\begin{picture}(6,3.5)
\put(-.15,0){\epsfig{file=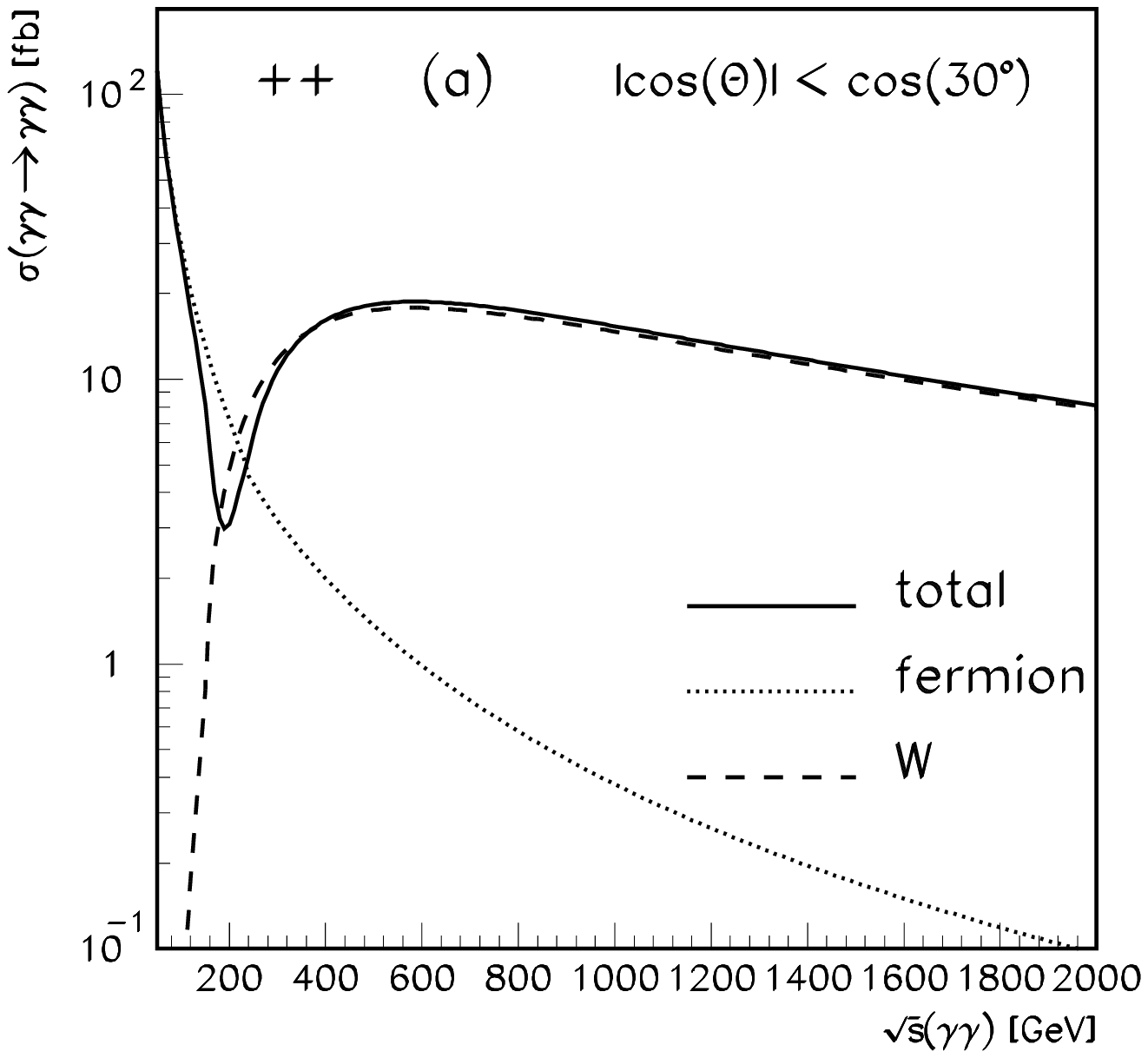,width=3.2in,height=3.5in}}
\put(2.85,0){\epsfig{file=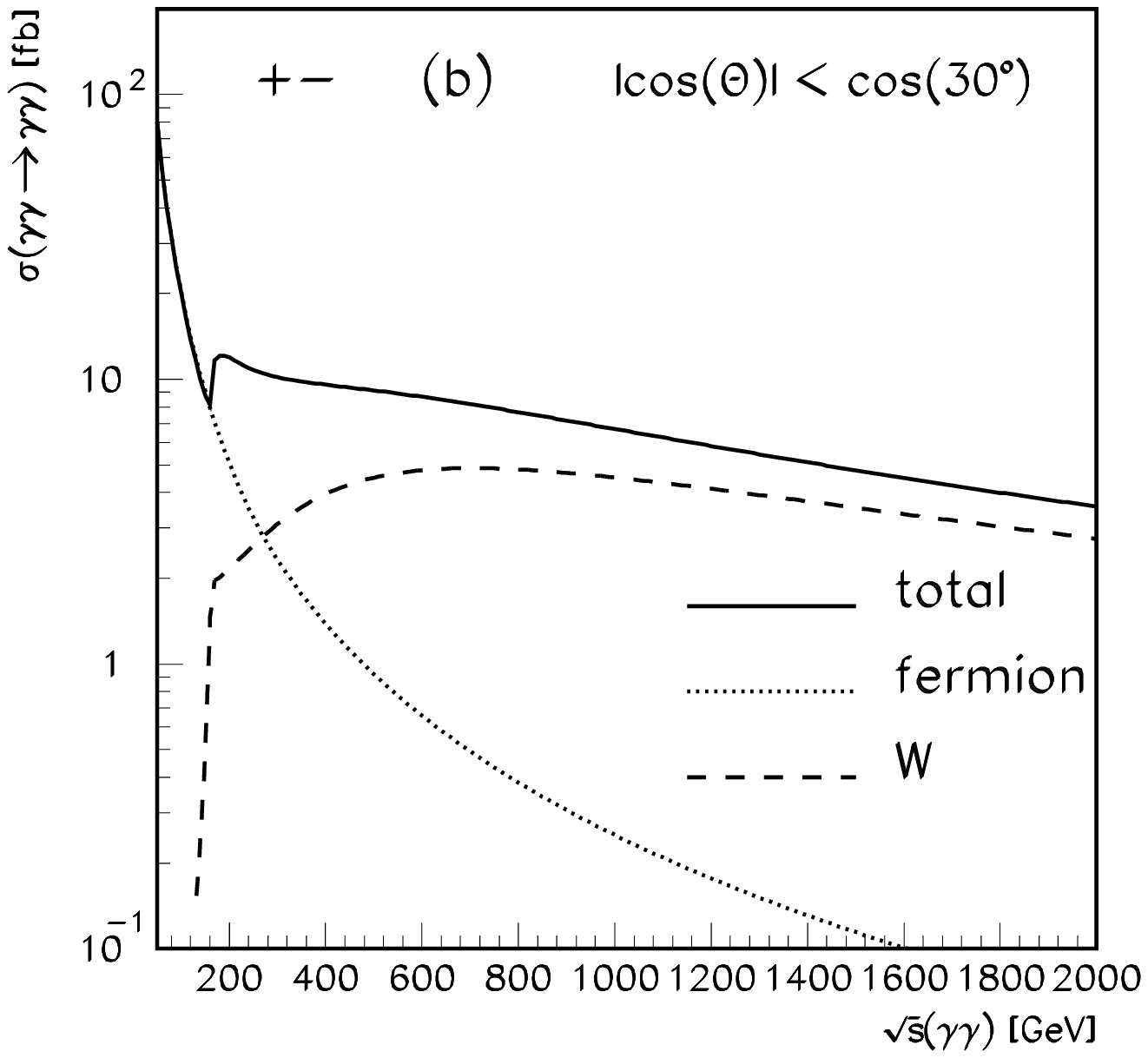,width=3.2in,height=3.5in}}
\end{picture}
\fcaption{Total cross section of photon-photon scattering in
monochromatic photon-photon collisions versus $\g\g$ c.m. energy for
different helicities of the incoming photons. Total cross section
(solid line) as well as $W$ boson loop contribution (dashed line) and
fermion loop contribution (dotted line) are shown.}
\end{figure}

\begin{figure}
\setlength{\unitlength}{1in}
\begin{picture}(6,3.5)
\put(-.15,0){\epsfig{file=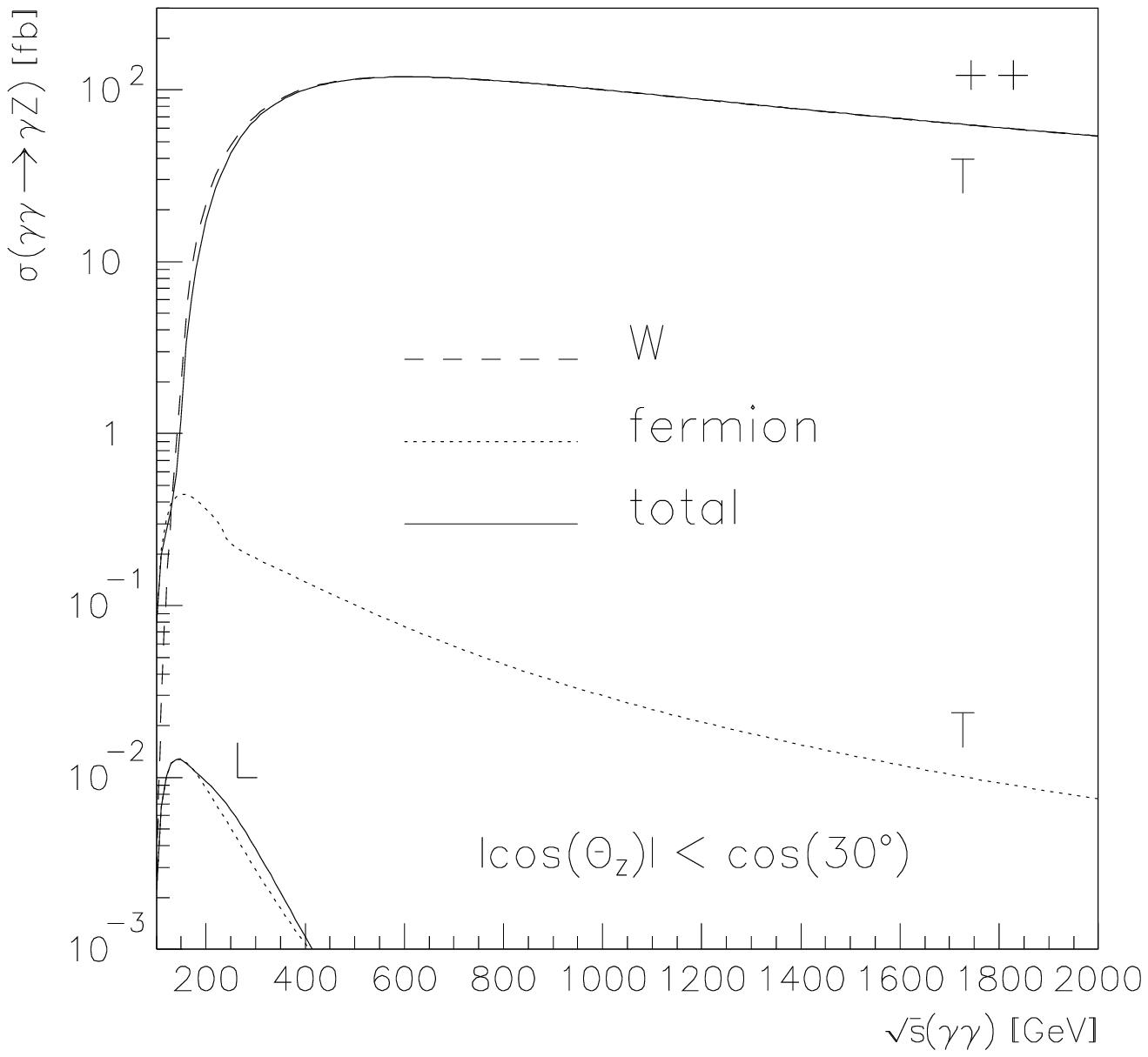,width=3.2in,height=3.5in}}
\put(2.85,0){\epsfig{file=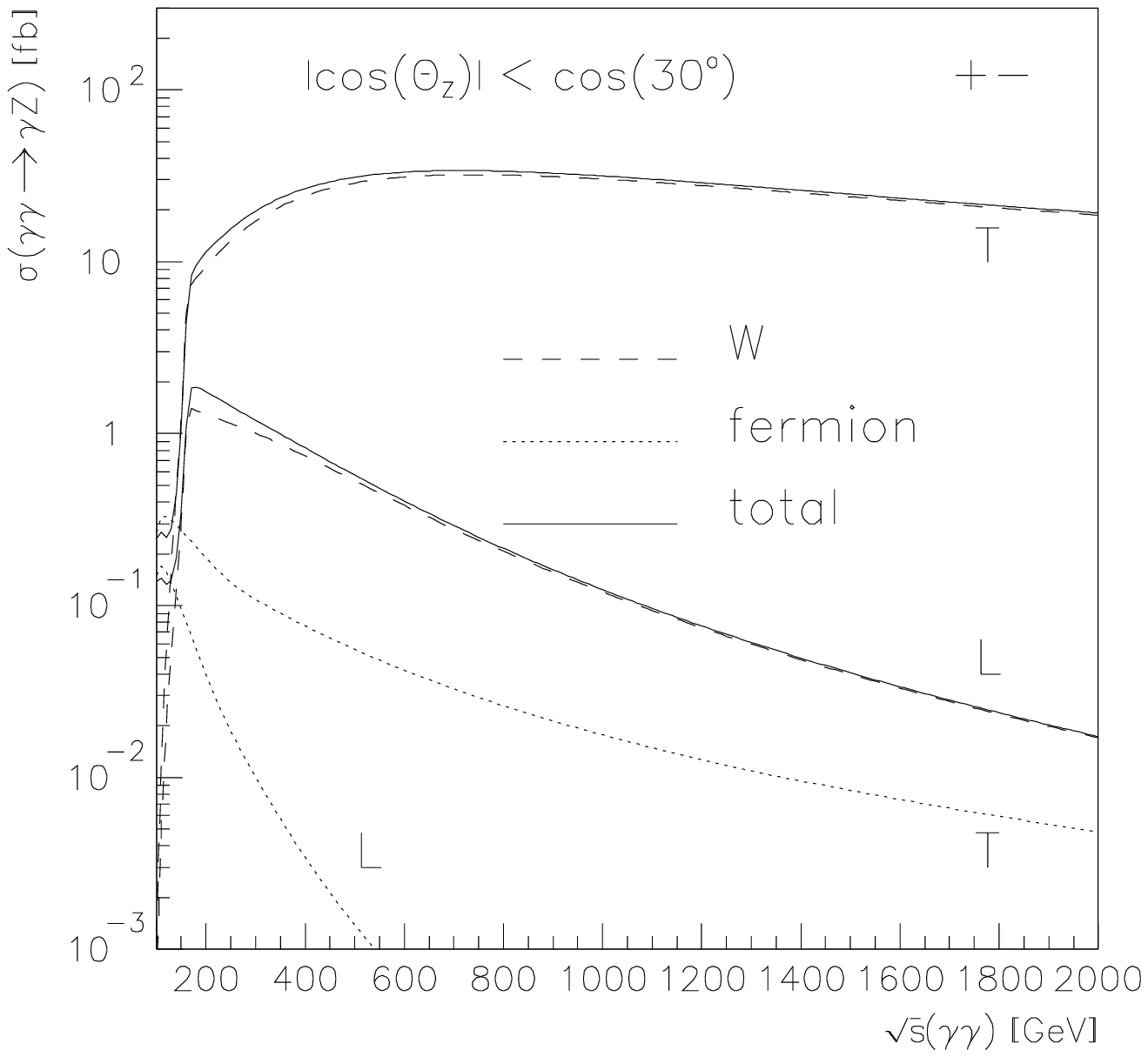,width=3.2in,height=3.5in}}
\end{picture}
\fcaption{Total cross section of $\g Z$ pair production in
monochromatic photon-photon collisions versus $\g\g$ c.m. energy for
different helicities of the incoming photons and final $Z$
boson. Total cross section (solid line) as well as $W$ boson loop
contribution (dashed line) and fermion loop contribution (dotted line)
are shown.}
\end{figure}

In fact, the measurement of $\g\g\to\g Z$ cross section is a
measurement of $Z\g\g\g$ coupling, which could be also measured in
three photon $Z$ decay.  However, is is well known, decay of the $Z$
boson into three photons via both fermion and $W$ boson loops in SM
has too small branching ratio (of the order of $3\cdot 10^{-10}$
\cite{zaaa}) to be observed at LEP experiments.  From the other side,
at PLC the $\gamma Z$ final state, which should be background free,
has the largest observable rate (if no light Higgs boson is present)
in comparison to $\gamma\gamma$ and $ZZ$ ({\it e.g.}, a three hundred
$\gamma Z$ pairs yearly can be produced at the Photon Collider
realized at the 500 GeV electron linear collider). So, even the unique
Standard Model $\g\g\g Z$ vertex can be measured in photon-photon
collisions.  Numerical values of the total cross section (with the
angular cuts imposed) are given in Table~3 \cite{AAAZ}.

\begin{table}[htb]
\tcaption{Event rates for $\g Z$ pair production at PLC}
\begin{center}
\begin{tabular}{|c||c|c|c|}\hline\hline
$\sqrt{s_{e^+e^-}}$	            & 300 GeV & 500 GeV &   1 TeV \\ \hline
$\sigma(\gamma\gamma\to\gamma Z_T)$ & 9.3 fb  & 32 fb   &   53 fb \\
$\sigma(\gamma\gamma\to\gamma Z_L)$ & 0.28 fb & 0.51 fb & 0.39 fb \\ 
\hline\hline
\end{tabular}
\end{center}
\end{table}

\end{document}